\documentclass[11pt,a4paper]{article}
\usepackage{amsmath,amsfonts,amssymb,amsthm,amstext,amscd,array,bbold}
\usepackage{fancyhdr}
\usepackage[dvips]{graphicx}
\usepackage{makeidx}
\usepackage{fancyhdr}
\usepackage[dvips]{graphicx}
\usepackage{makeidx}

\newcommand{\throw}{\longrightarrow}

\ifx\ll\undefined
  \newcommand{\ll}{\langle\langle}
\else
  \renewcommand{\ll}{\langle\langle}
\fi

\ifx\implies\undefined
  \newcommand{\implies}{\Longrightarrow}
\else
  \renewcommand{\implies}{\Longrightarrow}
\fi

\newcommand{\primitive}[1]{1 \otimes #1 + #1 \otimes 1}
\renewcommand{\mapsto}{\longmapsto}

\newcommand{\pref}[1]{(\ref{#1})}

\newcommand{\apply}{\triangleright}
\newcommand{\id}{\textrm{id}}

\newcommand{\hol}{{\rm hol}}

\def\ii{{\,{\rm i}\,}}
\def\dd{{\rm d}}

\def\mfg{{\mathfrak g}}

\newcommand{\eq}{\begin{equation}}
\newcommand{\eqend}{\end{equation}}
\newcommand{\eqa}{\begin{eqnarray}}
\newcommand{\nonueqa}{\begin{eqnarray*}}
\newcommand{\eqaend}{\end{eqnarray}}
\newcommand{\nonueqaend}{\end{eqnarray*}}

\newcommand{\bma}[1]{\begin{array}{#1}}
\newcommand{\ema}{\end{array}}
\newcommand{\bc}{\begin{center}}
\newcommand{\ec}{\end{center}}

\newcommand{\R}{\real}

\renewcommand{\thefootnote}{\fnsymbol{footnote}}
\newcommand{\newsection}{\setcounter{equation}{0}\section}

\def\appendix#1{\addtocounter{section}{1}\setcounter{equation}{0}
\renewcommand{\thesection}{\Alph{section}}
\section*{Appendix \thesection\protect\indent \parbox[t]{11.715cm} {#1}}
\addcontentsline{toc}{section}{Appendix \thesection\ \ \ #1} }

\newcommand{\complex}{{\mathbb C}} 
\newcommand{\nat}{{\mathbb N}} 
\newcommand{\real}{{\mathbb R}} 

\def\alg{{\mathcal A}}
\def\hil{{\mathcal H}}
\def\twist{{\mathcal F}}

\newif\ifold             \oldtrue

\newcommand{\Tr}[1]{\:{\rm Tr}\,#1}

\def\e{{\,\rm e}\,}

\def\be{\begin{equation}}
\def\ee{\end{equation}}
\def\bea{\begin{eqnarray}}
\def\eea{\end{eqnarray}}
\def\bd{\begin{displaymath}}
\def\ed{\end{displaymath}}

\newcommand{\beq}{\begin{eqnarray}}
\newcommand{\eeq}{\end{eqnarray}}

\makeatletter
\newdimen\normalarrayskip              
\newdimen\minarrayskip                 
\normalarrayskip\baselineskip
\minarrayskip\jot
\newif\ifold             \oldtrue            
\def\arraymode{\ifold\relax\else\displaystyle\fi} 
\def\@arrayskip{\ifold\baselineskip\z@\lineskip\z@
     \else
     \baselineskip\minarrayskip\lineskip2\minarrayskip\fi}
\def\@arrayclassz{\ifcase \@lastchclass \@acolampacol \or
\@ampacol \or \or \or \@addamp \or
   \@acolampacol \or \@firstampfalse \@acol \fi
\edef\@preamble{\@preamble
  \ifcase \@chnum
     \hfil$\relax\arraymode\@sharp$\hfil
     \or $\relax\arraymode\@sharp$\hfil
     \or \hfil$\relax\arraymode\@sharp$\fi}}
\def\@array[#1]#2{\setbox\@arstrutbox=\hbox{\vrule
     height\arraystretch \ht\strutbox
     depth\arraystretch \dp\strutbox
     width\z@}\@mkpream{#2}\edef\@preamble{\halign \noexpand\@halignto
\bgroup \tabskip\z@ \@arstrut \@preamble \tabskip\z@ \cr}%
\let\@startpbox\@@startpbox \let\@endpbox\@@endpbox
  \if #1t\vtop \else \if#1b\vbox \else \vcenter \fi\fi
  \bgroup \let\par\relax
  \let\@sharp##\let\protect\relax
  \@arrayskip\@preamble}
\makeatother

\begin{document}
\begin{titlepage}
\begin{flushright}

\baselineskip=12pt

HWM--07--6\\
EMPG--07--04\\
hep--th/0701273\\
\hfill{ }\\
January 2007
\end{flushright}

\begin{center}

\vspace{2cm}

\baselineskip=24pt

{\Large\bf Wilson Loops and Area-Preserving Diffeomorphisms \\ in
  Twisted Noncommutative Gauge Theory}

\baselineskip=14pt

\vspace{1cm}

{\bf Mauro Riccardi} and {\bf Richard J. Szabo}
\\[4mm]
{\it Department of Mathematics}\\ and\\ {\it Maxwell Institute for
Mathematical Sciences\\ Heriot-Watt University\\ Colin Maclaurin Building,
  Riccarton, Edinburgh EH14 4AS, U.K.}
\\{\tt mauro@ma.hw.ac.uk} , {\tt R.J.Szabo@ma.hw.ac.uk}
\\[40mm]

\end{center}

\begin{abstract}

\baselineskip=12pt

We use twist deformation techniques to analyse the behaviour under
area-preserving diffeomorphisms of quantum averages of Wilson
loops in Yang-Mills theory on the noncommutative
plane. We find that while the classical gauge theory is manifestly
twist covariant, the holonomy operators break the quantum
implementation of the twisted symmetry in the usual formal
definition of the twisted quantum field theory. These results are
deduced by analysing general criteria which guarantee twist invariance
of noncommutative quantum field theories. From this a number of general
results are also obtained, such as the twisted symplectic invariance of
noncommutative scalar quantum field theories with polynomial
interactions and the existence of a large class of holonomy operators
with both twisted gauge covariance and twisted symplectic invariance.

\end{abstract}

\end{titlepage}
\setcounter{page}{2}

\newpage

\renewcommand{\thefootnote}{\arabic{footnote}} \setcounter{footnote}{0}

\newsection{Introduction and Summary\label{Intro}}

Noncommutative gauge theory in two dimensions possesses many
interesting features that can be captured analytically and used to
shed light on generic features of noncommutative field theory. It is
an exactly solvable model whose partition function has been computed
explicitly as a semi-classical expansion over instantons on the
noncommutative torus in~\cite{szabo}--\cite{szaborev1}, as a sum over
fluxons on the noncommutative plane
in~\cite{Griguolo:2004jp,Griguolo:2001ce}, and also within its lattice
regularization in~\cite{Griguolo:2003kq,szaborev1}. The instanton
expansion has also been used to obtain exact analytic expressions for
quantum averages of open Wilson lines
in~\cite{Bassetto:2002kt,Griguolo:2001ce,Paniak:2003gn}. Exact Wilson
loop correlators, on the other hand, have remained
elusive. Perturbative computations on the noncommutative
plane~\cite{Ambjorn:2004ck,Bassetto:2005hn} reveal that Wilson loop
averages are not invariant under generic area-preserving
diffeomorphisms, in marked contrast to the commutative
case~\cite{Witten:1992xu}. The quantum gauge theory itself is only
invariant under the subgroup of global symplectic
transformations~\cite{Bassetto:2005hn} which is the largest spacetime
symmetry under which the Moyal product is
covariant~\cite{Gracia-Bondia:2006yj}. This breaks the topological
symmetry of the gauge theory making the exact semi-classical expansion
of Wilson loop correlators intractable. In~\cite{Cirafici:2005af} this
breakdown was interpreted nonperturbatively as the loss of covariance
under gauge Morita equivalence on the noncommutative torus. Other
aspects of Wilson loop correlators in two-dimensional noncommutative
Yang-Mills theory have been analysed using perturbation theory
in~\cite{Bassetto:2001is,Dorn:2003wi} and numerically
in~\cite{Bietenholz:2002ch}.

In parallel developments it has been realized how to reinstate
spacetime symmetries which are generically broken by noncommutativity
in noncommutative field theories~\cite{Aschieri:2005yw,CKNT1,Wess1}. The idea,
well-known from quantum group theory, is that the noncommutativity of
the algebra of functions $\alg$ should be compensated by a
noncommutative deformation of the bialgebra of diffeomorphisms acting
on $\alg$. For Moyal planes this is straightforward to do as the
star-product is defined in terms of an abelian Drinfeld twist element
and the required deformation is a twist deformation. In this paper we
will use twisted Hopf algebra techniques to investigate whether or not
the twisted form of area-preserving diffeomorphims can be used to
recover the full symplectic symmetry of two-dimensional noncommutative
Yang-Mills theory. Along the way we will add some general insights
into the properties of the much debated twisted noncommutative quantum
field theories. Twisted conformal transformations in two dimensions
have been analysed in~\cite{Lizzi:2006xi} while generic
two-dimensional twisted diffeomorphisms are discussed
in~\cite{Balachandran:2006qg}. Some parts of our analysis have obvious
generalizations to higher-dimensional Moyal planes and the behaviour
of generic noncommutative field theories under symplectomorphisms of
flat spacetime $\real^d$.

The origin of twisted spacetime symmetries can be understood in a
relatively straightforward way by proceeding abstractly as
follows~\cite{Balachandran:2006ib,Calmet:2004ii}. Let
$\alg=\alg_\theta$ be the associative algebra over $\complex$
generated by operators $\hat x^\mu$ obeying the Heisenberg commutation
relations
\beq
[\hat x^\mu,\hat x^\nu]=\ii\theta^{\mu\nu} \ .
\label{Heisenalg}\eeq
Let $\hat\partial_\mu\in{\rm der}(\alg)$ be linear derivations of
$\alg$ defined by
\beq
\big[\hat\partial_\mu\,,\,\hat x^\nu\big]=\delta_\mu{}^\nu \ , \quad
\big[\hat\partial_\mu\,,\,\hat\partial_\nu\big]=0 \ .
\label{linderivdef}\eeq
A generic infinitesimal diffeomorphism, implemented by a vector field
$\hat\xi=\xi^\mu(\hat x)\,\hat\partial_\mu$, is not an automorphism of
the algebra $\alg$ because it does not preserve the defining relations
(\ref{Heisenalg}). Consider now the new operators
\beq
x^\mu\doteq\hat x^\mu-\mbox{$\frac\ii2$}\,\theta^{\mu\nu}\,
\hat\partial_\nu \ , \qquad \partial_\mu\doteq\hat\partial_\mu
\label{newops}\eeq
which extend $\alg$ to the algebra spanned by ${\rm der}(\alg)$ and
generate a {\it commutative} algebra
\beq
[x^\mu,x^\nu]=0 \ , \qquad [\partial_\mu,x^\nu]=\delta_\mu{}^\nu \ ,
\qquad [\partial_\mu,\partial_\nu]=0 \ .
\label{commalg}\eeq
The meaning of the operators (\ref{newops}) is as follows. The
operators $\hat x^\mu$ generate the natural action of the algebra
$\alg$ on itself via multiplication from the left, whereas
$-\ii\theta^{\mu\nu}\,\hat\partial_\nu$ coincide with the commuting
adjoint action of $\hat x^\mu$ (using (\ref{Heisenalg})) and generate
the natural action of $\alg$ on itself via multiplication from the
right~\cite{Balachandran:2006ib}. This right $\alg$-module is
isomorphic to the opposite algebra $\alg'\cong\alg_{-\theta}$ which
defines the commutant of the algebra $\alg$ in the canonical
representation on itself. One has $\alg'\cong J\,\alg\,J^{-1}$ where
$J$ is the canonical Tomita involution~\cite{Lizzi:2001nd}. The
operators (\ref{newops}) are then the generators for the action of
$\alg$ on the Morita equivalence bimodule $\alg\otimes\alg'$.

The commutative bimodule algebra $\alg\otimes\alg'$ carries the
standard action of the diffeomorphism group, whose connected
components are generated by smooth vector fields
\beq
\xi=\xi^\mu(x)\,\partial_\mu \ .
\label{smoothvecfields}\eeq
Substituting (\ref{newops}) into (\ref{smoothvecfields}) using a
formal Taylor series expansion of the smooth functions $\xi^\mu(x)$
gives the corresponding operators acting on $\alg$ as
\bea
\hat\xi&\doteq&\xi^\mu\big(\mbox{$\hat x-\frac\ii2\,\theta\cdot
\hat\partial$}\,\big)\,\hat\partial_\mu \nonumber\\[4pt] &=&
\xi^\mu(\hat x)\,\hat\partial_\mu+\sum_{n=1}^\infty\,\left(-
\frac\ii2\right)^n\,\frac1{n!}\,\theta^{\mu_1\nu_1}\cdots
\theta^{\mu_n\nu_n}\,(\partial_{\mu_1}\cdots\partial_{\mu_n}
\xi^\mu)(\hat x)\,\hat\partial_{\nu_1}\cdots\hat\partial_{\nu_n}\,
\hat\partial_\mu \ .
\label{corropsdiff}\eea
Thus a vector field becomes a higher-order differential operator
acting on the noncommutative algebra $\alg$, and (\ref{corropsdiff})
is the usual expression for the twisted action of
diffeomorphisms~\cite{Aschieri:2005yw,Dimitrijevic:2004rf,Szabo:2006wx}.
For example, the generators of the Lorentz group
\beq
M^{\mu\nu}(x,\partial)=-\ii(x^\mu\,\partial^\nu-x^\nu\,\partial^\mu)
\label{Lorentzgenscomm}\eeq
act on $\alg$ as
\beq
\hat M^{\mu\nu}=M^{\mu\nu}\big(\hat x\,,\,\hat\partial\,
\big)-\mbox{$\frac\ii2$}\,
\big(\theta^{\mu\lambda}\,\hat\partial_\lambda\,\hat\partial^\nu-
\theta^{\nu\lambda}\,\hat\partial_\lambda\,\hat\partial^\mu\big) \ ,
\label{LorentzgensNC}\eeq
which is the usual action of twisted Lorentz
transformations~\cite{CKNT1,Wess1}.

The extra higher order terms imply that the diffeomorphisms do not act
as derivations of $\alg$ as the Leibnitz rule is deformed. This
deformation is exactly what is implemented in the twisted bialgebra of
diffeomorphisms. In particular, the noncommutativity tensor
$\theta^{\mu\nu}$ is in this way manifestly invariant under twisted
diffeomorphisms. On fields one uses the Weyl-Wigner
correspondence~\cite{Sz1} to implement the action of the
diffeomorphism group using
star-products~\cite{Aschieri:2005yw,Szabo:2006wx}.
In~\cite{Balachandran:2006ib} the usual commutative action on fields
is used instead, leading to a rather different class of twisted
noncommutative field theories than those considered here. A novel
interpretation of the Drinfeld twisting as deformed constraints in
canonical quantization of noncommutative field theory can be found
in~\cite{Rosenbaum:2006yu}.

When transcribed to quantum field theory, the relative simplicity of
the implementation above appears to be in agreement with
suggestions~\cite{Fiore:2007vg,Zahn:2006wt} that there are no
non-trivial observable consequences of the twisted Poincar\'e symmetry
in noncommutative field theory. However, most of the debates
surrounding such issues deal only with scalar fields and it is not
presently clear what the situation is for gauge theories. To try to
understand this better, in the following we will study the action of
these twisted diffeomorphisms on the observables of two-dimensional
noncommutative Yang-Mills theory. We will first study the action of
the twisted diffeomorphism group on generic noncommutative quantum
field theories in two dimensions. We will find, quite generally, that
classical invariance already truncates to the subgroup of
area-preserving diffeomorphisms, as expected but in a rather
non-trivial way. In particular, generic scalar quantum field
theories with polynomial interactions are twist invariant under
symplectic transformations. When
applied to gauge theory in two dimensions one can immediately infer
from this the twisted symplectic invariance of the classical
noncommutative Yang-Mills action and all observables. This ties in
nicely with the geometric interpretation of noncommutative gauge
symmetries as ``deformed'' symplectomorphisms of flat
spacetime~\cite{Lizzi:2001nd}.

We point out the existence of two natural sets of holonomy observables
in noncommutative gauge theory, one associated with the ordinary
tensor algebra of fields and the other with the braided tensor
algebra. Both classes of operators are manifestly covariant under both
twisted gauge
transformations~\cite{Aschieri:2006ye,Vassilevich:2006tc} and twisted
area-preserving diffeomorphisms. The latter class defines the usual
noncommutative Wilson loops which are in addition invariant under
star-gauge transformations~\cite{DougNek1,Sz1}. Although classically
invariant under twisted area-preserving diffeomorphisms, the quantum
averages of these holonomies break the twisted spacetime symmetry. The
twisted symmetry is not anomalous but is instead broken by the quantum
implementation of Wilson loop operators in star-gauge invariant
correlation functions. Thus only some non-standard definition of the
twisted quantum gauge theory can allow for full invariance under the
symplectic symmetry. This provides a much more general, explicit and
nonperturbative description for the loss of topological symmetry than
those given
in~\cite{Ambjorn:2004ck,Bassetto:2005hn,Cirafici:2005af}. On the other
hand, the twisted symmetry of the quantum observables under global
symplectic transformations follows rather directly from the structure
of the twisted bialgebra of area-preserving diffeomorphisms.

The organisation of the remainder of this paper is as follows. We
begin in Section~\ref{TwistSym} with a general description of twisted
symmetries and their specialization to the diffeomorphism group of
spacetime, describing in particular the twisted bialgebra of
area-preserving diffeomorphisms. In Section~\ref{TwistNCFT} we
describe the generic implementation of twisted diffeomorphisms as
symmetries of the quantum effective action in noncommutative field
theories, showing in particular that noncommutative scalar quantum
field theory is twist invariant under symplectic
transformations. Finally, in Section~\ref{WilsonTNGT} we detail the
construction of a broad set of holonomy observables in twisted
noncommutative gauge theory, and study in detail the behaviour of
star-Wilson loops under twisted area-preserving diffeomorphisms.

\newsection{Twisted Spacetime Symmetries\label{TwistSym}}

In this section we will discuss some basic general aspects of twisted
symmetry groups of associative algebras in the context of Hopf
algebras and Drinfeld
twists~\cite{Drinfeld:1989st,Oeckl:2000eg,Resh1} (see~\cite{QG1} for
the necessary background on quantum groups). We will then
specialize the discussion,
following~\cite{Aschieri:2005zs,Aschieri:2005yw}, to the case of
diffeomorphisms acting on fields on the Moyal plane. In particular, we
describe the twisted bialgebra of area-preserving diffeomorphisms of
$\R^2$, which will be the most pertinent example in the subsequent
sections.

\subsection{Twist Deformations of Group Actions\label{sec:twistdelta}}

Let $\alg$ be an associative algebra over $\complex$ with multiplication map
$\mu_0:\alg\otimes\alg\to\alg$. Let $G\subset{\rm Aut}(\alg)$ be a
group of symmetries of $\alg$ acting by automorphisms. Let
$\hil=\complex G$ be the group algebra of $G$. We can make $\hil$ into
a Hopf algebra whose underlying bialgebra structure is specified by
the cocommutative primitive coproduct which is the homomorphism
$\Delta_0:\hil\to\hil\otimes\hil$ defined by $\Delta_0(g)=g\otimes g$
on generators $g\in G$ and $\Delta_0(\id)=\id\otimes\id$. The coproduct
implements the action of $G$ on the tensor product $\alg\otimes\alg$
with the fundamental compatibility condition
\begin{equation}
g\circ \mu_0 (a\otimes b) =  \mu_0 \circ \Delta_0(g)(a\otimes b)
\label{eqn:fundid}
\end{equation}
for all $g\in\hil$ and $a,b\in\alg$. This is just the expected
covariance condition on the multiplication in $\alg$, and it is a
necessary and sufficient condition for the group $G$ to act on the
algebra $\alg$.

We introduce an operator $\twist\in\hil\otimes\hil$, called a twist
element~\cite{Drinfeld:1989st,Resh1}, which acts on the tensor product
\beq
\mathcal F \,:\, \mathcal A \otimes \mathcal A ~
\longrightarrow~ \mathcal A \otimes \mathcal A 
\eeq
such that the pair
\beq
\big(\mathcal A\,,\,\mu \doteq \mu_0\circ\mathcal F\big)
\label{twistalg}\eeq
is still an associative algebra. If the original multiplication map
$\mu_0$ is commutative, then generically the twisted product $\mu$ is
no longer commutative. We will also write the twisted product using
the star product notation $\mu(a\otimes b) \doteq a\star
b$. Compatibility of the action of $G$ on the twist deformed algebra
(\ref{twistalg}) requires twisting of the bialgebra structure on
$\hil$ as well. The condition \pref{eqn:fundid} then becomes
\begin{equation}
g\circ \mu (a\otimes b) =  \mu \circ \Delta(g)(a\otimes b)
\label{eqn:twistfundid}
\end{equation}
for all $g\in\hil$ and $a,b\in\alg$. The generically
non-cocommutative twisted coproduct defined by the similarity
transformation
\beq
\Delta \,:\,\mathcal H~\longrightarrow~
\mathcal H\otimes\mathcal H \ , \qquad
\Delta \doteq \mathcal F^{-1} \circ \Delta_0 \circ \mathcal F
\eeq
makes $\hil$ into a triangular Hopf algebra. All of this requires
$\twist$ to be an invertible, co-unital two-cocycle of
the Hopf algebra $\hil$. In the star product notation we will write
the left-hand side of (\ref{eqn:twistfundid}) as $g\circ \mu (a\otimes
b)\doteq g\apply(a\star b)$.

In many cases of interest the algebra $\alg$ will carry generic
products of arbitrary representations of the Hopf algebra $\hil$. If
an element $a\in\mathcal A$ transforms in a representation $R$ of the
symmetry group $G$, then the corresponding action of an automorphism
$g\in\mathcal H$ is denoted
\beq
g\,:\, \mathcal A ~\throw~ \mathcal A \ , \qquad  a ~\mapsto~
 g^{(R)}(a) \ .
\eeq
On tensor products $a_1\otimes a_2$, with $a_i\in\alg$ transforming in
the representation $R_i$, the operator $g$ acts via the coproduct as
\beq
g \triangleright (a_1\otimes a_2) = \Delta(g)(a_1\otimes a_2) =
\big(g^{(R_1)}_{:1}\otimes g^{(R_2)}_{:2}\big)(a_1\otimes a_2) \ .
\eeq
Here we have used the Sweedler notation $\Delta(g)=g_{:1}\otimes
g_{:2}$ (with a sum over $g_{:1}$ and $g_{:2}$ understood).

Let us now {\em assume} that the twist leaves each factor of the
tensor product $a_1\otimes a_2$ in its same representation. We can
then write the compatibility condition \pref{eqn:twistfundid} for
$g\in G$ as
\begin{equation}
\begin{split}
g \apply (a_1\star a_2) = & ~\mu_0 \circ \Delta_0(g)\circ\mathcal F
(a_1\otimes a_2) \\[4pt] = & ~\mu_0 \circ \big(g^{(R_1)}\otimes g^{(R_2)}
\big) \circ\mathcal F (a_1\otimes a_2) \\[4pt]
= & ~\mu_0 \circ \big(g^{(R_1)}\otimes g^{(R_2)}\big) \circ\mathcal F
\circ \big(g^{(R_1)}\otimes g^{(R_2)}\big)^{-1} \circ \big(g^{(R_1)}
\otimes g^{(R_2)}\big)(a_1\otimes a_2) \ .
\end{split}
\label{eqn:prodtransf}
\end{equation}
It follows that the group $G$ acts on the twist operator $\mathcal F$
itself through the similarity transformations
\beq
g\,:\, \mathcal F ~\mapsto~ g\apply\mathcal F = 
\big(g^{(R_1)}\otimes g^{(R_2)}\big) \circ\mathcal F \circ 
\big(g^{(R_1)}\otimes g^{(R_2)}\big)^{-1} \ .
\label{twisttransf}\eeq

In our subsequent applications we will be interested in the
infinitesimal version of this construction for the case when $G$
is a continuous group of symmetries with Lie algebra $\mfg$. In that
case we take $\hil=U(\mfg)$ to be the universal enveloping algebra of
$\mfg$, made into a Hopf algebra with bialgebra structure specified by
the primitive coproduct $\Delta_0(X)=X\otimes\id+\id\otimes X$ on generators
$X\in\mfg$. Then the compatibility condition (\ref{eqn:fundid})
implements the usual Leibnitz rule for the action of the symmetry
algebra $\mfg$ on $(\alg,\mu_0)$, while the twist deformed Leibnitz
rule (\ref{eqn:twistfundid}) allows for a representation of $\mfg$ on
(\ref{twistalg}). The corresponding action of the Lie algebra on the
twist is given by
\begin{equation}
\begin{split}
X\,:\, \mathcal F ~\mapsto~ X\apply\mathcal F = &~ \big[
X^{(R_1)}_{:1}\otimes X^{(R_2)}_{:2} \,,\, \mathcal F \big] \\[4pt]
= & ~\big[X^{(R_1)}\otimes \id^{(R_2)} + \id^{(R_1)}\otimes X^{(R_2)} 
\,,\, \mathcal F \big] \ .
\end{split}
\label{twisttransfinf}\end{equation}

This machinery can also be applied if the twist element does {\em not}
leave each factor of $a_1\otimes a_2$ in its own representation. In
the general case we can write the compatibility condition
\pref{eqn:prodtransf} in a more generic way as
\beq
g \apply (a_1\star a_2) = \mu_0 \circ \big(g^{(R_1)}\otimes g^{(R_2)}
\big) \circ\mathcal F \circ \big(g^{(L_1)}\otimes g^{(L_2)}
\big)^{-1} \circ \big(g^{(L_1)}\otimes g^{(L_2)}\big)(a_1\otimes a_2)
\ ,
\eeq
where the twist $\twist$ maps the $R_i$ representation of $G$ into
$L_i$. This modifies the similarity transformation (\ref{twisttransf})
of the twist operator to
\beq
g\,:\, \mathcal F ~\mapsto~ g\apply\mathcal F = 
\big(g^{(R_1)}\otimes g^{(R_2)}\big) \circ\mathcal F \circ 
\big(g^{(L_1)}\otimes g^{(L_2)}\big)^{-1} \ .
\eeq

\subsection{Braided Tensor Calculus\label{TwistDiffs}}

The generic situation of interest for noncommutative field theory is
when $\alg$ is taken to be the algebra of smooth functions on a
manifold $\mathcal M$ (with appropriate boundary conditions when
$\mathcal M$ is non-compact), with pointwise multiplication $\mu_0$,
and with $G={\rm Diff}(\mathcal{M})$ the group of diffeomorphisms of
$\mathcal M$. In this case the algebra $\mathcal A$ will contain
arbitrary products of representations of the corresponding Hopf
algebra $\mathcal H$ determined by tensorial rank, and one is in the
situation described in Section~\ref{sec:twistdelta} above. In this
paper we are primarily interested in the case where
$\mathcal{M}=\real^2$ and the twist element is the bidifferential
operator
\beq
\mathcal F =\exp\left(\mbox{$\frac{\ii\theta}{2}$}\,
\left(\partial_1\otimes\partial_2-\partial_2\otimes\partial_1\right)
\right) = \sum_{k=0}^\infty\, \frac{(\ii\theta/2)^k}{k!}\,
\left(\partial_1\otimes\partial_2-
\partial_2\otimes\partial_1\right)^k
\label{Moyaltwist}\eeq
with $\theta\in\real$ constant. This defines an abelian Drinfeld twist
and the corresponding star product is the Moyal product on the algebra
of functions on $\real^2$. Many of the following results hold
analogously for higher-dimensional Moyal planes.

The connected components of the diffeomorphism group are generated by
smooth vector fields $\xi=\xi^\mu(x)~\partial_\mu$. Their
infinitesimal action on $\alg$ is denoted $\delta_\xi$, and the
elements of $\hil$ act on $\alg$ by the natural extension of the Lie
derivative. Since the twist operator (\ref{Moyaltwist}) is constructed
from translation generators, i.e. constant diffeomorphisms, each
factor of any tensor product $f_1\otimes f_2$ of two fields remains in
its own representation under the action of $\mathcal F$. This follows
from the identity
\beq
\left[\partial_\mu,\delta_\xi\right] = \delta_{\partial_\mu\xi} \ .
\eeq
Then the transformation (\ref{twisttransfinf}) of the twist in
the case at hand reads
\begin{equation}
\begin{split}
\delta_\xi\apply\mathcal F = & - \mathcal F \circ \sum_{k=1}^\infty\, 
\frac{({-\ii\theta}/{2})^k}{k!}\,\overbrace{\big[
\partial_1\otimes\partial_2-\partial_2\otimes\partial_1\,,\,\dots\,,\,
\big[\partial_1\otimes\partial_2-\partial_2\otimes\partial_1}^k,
\Delta_0(\delta_\xi)\big]\dots\big] \\[4pt]
= & - \mathcal F \circ \sum_{k=1}^\infty\, \frac{({-\ii\theta}/
{2})^k}{k!}~\sum_{l=0}^k\,\binom{k}{l}\,(-1)^l\,
\bigg(\delta^{(R_1)}_{\partial_1^{k-l}\partial_2^l\xi}\otimes\partial_2^{k-l}
\partial_1^l+ (-1)^k\, \partial_2^{k-l}\partial_1^l \otimes
\delta^{(R_2)}_{\partial_1^{k-l}\partial_2^l\xi}\bigg) \ .
\end{split}
\label{Moyaltwisttransf}\end{equation}
It follows that the transformation of the star product of two fields
under an infinitesimal diffeomorphism of $\real^2$ is given by
\begin{equation}
\begin{split}
\delta_\xi\apply (f_1\star f_2) = -\mu\Big(~
\mbox{$\sum\limits_{k=0}^\infty\, \frac{({-\ii\theta}/{2})^k}
{k!}~\sum\limits_{l=0}^k\,\binom{k}{l}$}\,(-1)^l\, \big(&
\delta^{(R_1)}_{\partial_1^{k-l}\partial_2^l\xi}f_1\otimes\partial_2^{k-l}
\partial_1^lf_2 \\ & + (-1)^k \,\partial_2^{k-l}\partial_1^l f_1 \otimes
\delta^{(R_2)}_{\partial_1^{k-l}\partial_2^l\xi}f_2\big)\Big) \ .
\end{split}
\label{eqn:gentransrule}
\end{equation}

Let us work out some examples of the transformation rule
(\ref{eqn:gentransrule}) which will be used in the sequel. If both
$f_1$ and $f_2$ are scalar fields in $\mathcal A$, then
$\delta_\xi^{(R_1)} =\delta_\xi^{(R_2)}$ is the differential operator
$-\xi^\mu~\partial_\mu$ and one has
\begin{align}
\delta_\xi\apply (f_1\star f_2) 
= - \mu\Big(~\mbox{$\sum\limits_{k=0}^\infty\,
  \frac{({-\ii\theta}/{2})^k}{k!}~\sum\limits_{l=0}^k\,\binom{k}{l}\,$}&
(-1)^l\, \big(  (\partial_1^{k-l}\partial_2^l\xi^\mu)\,\partial_\mu f_1
\otimes\partial_2^{k-l}\partial_1^l f_2  \nonumber\\
\phantom{.} & + (-1)^k\, \partial_2^{k-l}\partial_1^l f_1 
\otimes (\partial_1^{k-l}\partial_2^l\xi^\mu)\,\partial_\mu
f_2\big)\Big) \ .
\end{align}
This can be seen to be equal to the expression
\begin{equation}
\begin{split}
&\mu_0\big(-(\xi^\mu~\partial_\mu \otimes \id + \id \otimes \xi^\mu~
\partial_\mu )\circ\mathcal F(f_1\otimes f_2)\big) \\
&\qquad\qquad ~=~ - \sum_{k=0}^\infty\, \frac{({\ii\theta}/
{2})^k}{k!}~\sum_{l=0}^k\,\binom{k}{l}\,(-1)^l\, 
\Big((\xi^\mu\, \partial_1^{k-l}\partial_2^l \partial_\mu f_1)
\,(\partial_2^{k-l}\partial_1^lf_2) \\ & \qquad\qquad\qquad\qquad
\qquad\qquad\qquad\qquad\qquad\qquad + (-1)^k\, (\partial_2^{k-l}
\partial_1^l f_1) \,(\xi^\mu\,\partial_1^{k-l}\partial_2^l\partial_\mu
f_2)\Big) \ .
\end{split}
\end{equation}
Hence if both $f_1$ and $f_2$ are scalar fields, then their
star-product $f_1\star f_2$ also transforms as a scalar field,
\beq
\delta_\xi\apply (f_1\star f_2) = -\mu_0\big((\xi^\mu~
\partial_\mu \otimes \id + \id \otimes \xi^\mu~
\partial_\mu)\circ\mathcal F(f_1\otimes f_2)\big) = -\xi^\mu\,
\partial_\mu (f_1\star f_2) \ .
\eeq

If instead we take the product of two tensor fields, then we have to
account for the transformations with respect to contractions of the
indices as well. From \pref{eqn:gentransrule} we can work out as an
example the case of two rank one tensor fields $f_1=V_\mu$ and
$f_2=W^\nu$. Since
\bea
\delta_\xi^{(R_1)} V_\mu & = & -\xi^\sigma\,\partial_\sigma V_\mu - (
\partial_\mu\xi^\sigma)\, V_\sigma \ , \nonumber\\[4pt]
\delta_\xi^{(R_2)} W^\mu & = & -\xi^\sigma\,\partial_\sigma W^\mu + (
\partial_\sigma\xi^\mu)\, W^\sigma \ ,
\label{deltaxirank1}\eea
we easily find that
\begin{equation}
\begin{split}
\delta_\xi\apply (V_\mu\star W^\nu) = & -\xi^\sigma\,\partial_\sigma
(V_\mu\star W^\nu) \\ & + \mu\Big(~\mbox{$ \sum\limits_{n=0}^\infty\,
  \frac{({-\ii\theta}/{2})^n}{n!}~\sum\limits_{l=0}^n\,
\binom{n}{l}$}\,(-1)^l\,\Big[\big(-(\partial_1^{n-l}\partial_2^l
\partial_\mu\xi^\sigma) \,V_\sigma\big)\otimes 
\big(\partial_2^{n-l}\partial_1^l W^\nu\big) \\
&\hspace{6cm}+ \big(\partial_2^{n-l}\partial_1^l V_\mu\big)\otimes 
\big((\partial_1^{n-l}\partial_2^l\partial_\sigma\xi^\mu)\,
 W^\sigma\big)\Big]\Big) \\[4pt]
= & -\xi^\sigma\,\partial_\sigma (V_\mu\star W^\nu) \\
& + \mu_0\big( \mathcal F\circ(V_\sigma\otimes 1)\circ\mathcal F^{-1}
(-\partial_\mu\xi^\sigma\otimes W^\nu) + \mathcal F\circ
(1\otimes W^\sigma)\circ\mathcal F^{-1}(V_\mu\otimes
-\partial_\sigma\xi^\nu)\big) \ .
\end{split}
\end{equation}
Using the similarity transformations
\bea
\mathcal F\circ(V_\sigma\otimes 1\big)\circ\mathcal F^{-1} &=& 
\sum_{n=1}^\infty\,\frac{({\ii\theta}/{2})^n}{n!}~
\sum_{l=0}^n\,\binom{n}{l}\,(-1)^l\,\big(\partial_1^{n-l}\partial_2^l
V_\sigma\big)\otimes \partial_2^{n-l}\partial_1^l \ , \nonumber\\[4pt]
\mathcal F\circ(1\otimes W^\sigma)\circ\mathcal F^{-1}  &=& 
\sum_{n=0}^\infty\,\frac{({\ii\theta}/{2})^n}{n!}~
\sum_{l=0}^n\,\binom{n}{l}\,(-1)^l\,\partial_1^{n-l}\partial_2^l 
\otimes \big(\partial_2^{n-l}\partial_1^l W^\sigma\big) \ ,
\eea
we obtain
\begin{equation}
\begin{split}
\delta_\xi\apply (V_\mu\star W^\nu) = & -\xi^\sigma\,\partial_\sigma
(V_\mu\star
W^\nu) \\ & + \mu_0\big( (-\partial_\mu\xi^\sigma\otimes 1)\circ
\mathcal F(V_\sigma\otimes W^\nu)\big) + \mu_0\big(
(1\otimes\partial_\sigma\xi^\nu)\circ\mathcal F(V_\mu\otimes
W^\sigma)\big)
\end{split}
\end{equation}
which is the usual rule of tensor calculus for the transformation of a
tensor field of rank $(1,1)$.

In particular, the contraction $V_\mu\star W^\mu$ transforms
expectedly as a scalar field,
\bea
\delta_\xi\apply (V_\mu\star W^\mu) &=& -\xi^\sigma\,
\partial_\sigma (V_\mu\star W^\mu) + \mu_0\big( (1\otimes\partial_\mu
\xi^\sigma-\partial_\mu\xi^\sigma\otimes 1)\circ
\mathcal F(V_\sigma\otimes W^\mu)\big) \nonumber\\[4pt]
&=& -\xi^\sigma\,
\partial_\sigma (V_\mu\star W^\mu) \ ,
\label{deltacontr}\eeq
because in the second term the derivative operators of $\mathcal F$
do not act on $\xi$. Note that if the vector field $\xi$ generates a
linear affine transformation in
$\mathfrak{gl}(2,\real)\rtimes\real^2$, given by
\beq
\xi^\sigma(x) = L^\sigma\!_\mu \,x^\mu + a^\sigma
\label{affine}\eeq
with constant tensors $L^\sigma\!_\mu$ and $a^\sigma$, then one has
\beq
1\otimes\partial_\mu\xi^\sigma = L^\sigma\!_\mu (1\otimes 1) = 
\partial_\mu\xi^\sigma\otimes 1 \ .
\eeq
This automatically guarantees the scalar property of the contraction
(\ref{deltacontr}) under global diffeomorphisms.

The general result~\cite{Aschieri:2005zs,Aschieri:2005yw} is that
covariant expressions in ordinary (untwisted, commutative) tensor
calculus are still covariant after the twist deformation. In
Section~\ref{WilsonTNGT} we will use this feature to
infer the classical invariance of two-dimensional noncommutative
Yang-Mills theory under twisted area-preserving diffeomorphisms. Note
that since the compatibility equation \pref{eqn:twistfundid} is
equivalent to the untwisted one in \pref{eqn:fundid}, the tensor calculus
constructed by using the standard pointwise multiplication $\mu_0$ is
still covariant as well under {\em untwisted} diffeomorphisms.

\subsection{Twisted Area-Preserving Diffeomorphisms\label{APD}}

In order to gain a better understanding of the formalism above, we now
display explicitly the twisted bialgebra structure of the Lie algebra
of area-preserving diffeomorphisms of $\real^2$, which will be our
prominent example of a twisted symmetry in this paper. These are the
symplectic transformations preserving the area form $\dd^2x=\dd
x^1\wedge\dd x^2$. Let us denote the generators by $\mathfrak{L}_{m,n}$,
$m,n\in\nat$. We select the representation on $\alg$ given by the
differential operators
\beq
\mathfrak{L}_{n,m} = m \,x_1^n \,x_2^{m-1}~ \partial_1- n\,
x_1^{n-1}\, x_2^m ~
\partial_2 \ .
\eeq
The elementary Lie brackets are those of the
infinite-dimensional $W_{1+\infty}$ algebra
\beq
[\mathfrak{L}_{m,n},\mathfrak{L}_{p,q}]=(n\,p-m\,q)~
\mathfrak{L}_{m+p-1,n+q-1} \ .
\eeq
Then the twisted coproduct of these generators can be
straightforwardly worked out to be
\bea
\Delta(\mathfrak{L}_{n,m}) &=&\mathfrak{L}_{n,m}\otimes1+1\otimes
\mathfrak{L}_{n,m}\\ &&
+ \sum_{k=1}^{n+m-1}\,({-\ii\theta}/{2})^k~
\sum_{l=0}^k\,\binom{n}{k-l}\,\binom{m}{l}\,(-1)^l\,
\Big[(-1)^k~\mathfrak{L}_{n-k+l,m-l}\otimes\big(\partial_1^{k-l}
\partial_2^l\big) \nonumber\\ && \nonumber \hspace{8cm}
+ \big(\partial_1^{k-l}\partial_2^l\big)\otimes
\mathfrak{L}_{n-k+l,m-l}\Big] \ .
\label{eqn:genercop}
\eea

In particular, for the generators $\mathfrak{L}_{m,n}$ with $n+m\leq2$
one finds
\begin{eqnarray}
\Delta(\mathfrak{L}_{0,1}) & = & \Delta(\partial_1) ~=~
\primitive{\partial_1} \ , \nonumber\\[4pt]
\Delta(\mathfrak{L}_{1,0}) & = & \Delta(-\partial_2) ~=~
\primitive{(-\partial_2)} \ , \nonumber\\[4pt]
\Delta(\mathfrak{L}_{1,1}) & = & \Delta(x_1~\partial_1-x_2~
\partial_2) ~=~ 
\primitive{(x_1~\partial_1-x_2~\partial_2)} \ , \nonumber\\[4pt]
\Delta(\mathfrak{L}_{2,0}) & = & \Delta(- 2 x_1~\partial_2) ~=~ 
\primitive{(- 2 x_1~\partial_2)} \ , \nonumber\\[4pt]
\Delta(\mathfrak{L}_{0,2}) & = & \Delta(2 x_2~\partial_1) ~=~ 
\primitive{(2 x_2~ \partial_1)} \ .
\end{eqnarray}
Thus these operators are all primitive with respect to the twisted
coproduct $\Delta$, and hence the tensor products of their
representation is unchanged. They generate a subalgebra of
$W_{1+\infty}$ isomorphic to $\mathfrak{sl}(2,\mathbb
R)\rtimes\real^2$, comprising {\em global} diffeomorphisms which
generate linear unimodular affine transformations with $L^\mu{}_\mu=0$
in (\ref{affine}). We can further easily check from
\pref{eqn:genercop} that these are the only generators whose primitive
coproducts are unchanged by the twist, since higher order terms in
$\theta$ can only cancel when {\em simultaneously} $k\le 1$ and
$k-l\le 1$, due to the fact that the generators $\mathfrak{L}_{m,n}$
are linear in derivative operators. Therefore the Lie algebra
$\mathfrak{sl}(2,\mathbb R)\rtimes\real^2$ of global area-preserving
diffeomorphisms is the largest subalgebra of $W_{1+\infty}$ whose
representations are unaffected by the twist. This fact will be crucial
in determining under which symmetry the physical quantum observables of
noncommutative Yang-Mills theory have twisted invariance. It provides
an alternative way to understand the known covariance of the Moyal
product~\cite{Gracia-Bondia:2006yj} and of noncommutative gauge
theory~\cite{Bassetto:2005hn} under linear affine transformations.

\newsection{Spacetime Symmetries of Twisted Noncommutative Field\\
  Theory\label{TwistNCFT}}

In this section we will perform a general analysis of the invariance
of two-dimensional noncommutative field theory under the twisted
spacetime symmetries of the previous section. We adapt the point of
view that the twist deformation realizes diffeomorphisms as potential
internal symmetries of noncommutative field
theory~\cite{Lizzi:2001nd,Sz1}, so that some of the ensuing statements
also apply to other classes of twisted symmetries. We
will find that, generically, already at the classical level
the twisted invariance truncates to the area-preserving diffeomorphisms of
Section~\ref{APD}. However, this truncation does {\it not} arise from
setting the Jacobian of a map equal to one in order to preserve the area
form $\dd^2x$ in the action, as the twist does not act on any integration
measure. At the quantum level, we will interpret the
potential loss of twisted symmetry as anomalous behaviour of the field
theory. In the next section these general considerations will be
applied to Wilson loop correlators in noncommutative Yang-Mills
theory.

\subsection{Symmetries and Ward Identities}

We begin by recalling how to implement (spacetime) symmetries in
quantum field theory. Consider a (noncommutative) field theory on
$\real^2$ with fields $\Phi_i$ and action functional $S[\Phi]$. The
symbol $\Phi_i$ in general collectively denotes all fundamental fields
such as matter and gauge fields, as well as any auxilliary and ghost
fields, and $S[\Phi]$ may generally include Lagrange multiplier and
gauge-fixing terms. Under a symmetry transformation of the field
theory (for instance a diffeomorphism of spacetime), the infinitesimal
variation of the fields is denoted
\beq
\Phi_i ~\mapsto~ \Phi_i + \delta\Phi_i \ .
\label{deltaPhi}\eeq
We wish to implement this transformation as a symmetry of the quantum
correlation functions of a set of operators
$\mathcal{O}_1(\Phi),\dots,\mathcal{O}_n(\Phi)$ in the quantum field
theory. In path integral quantization, this roughly leads to the
identification
\beq
\int\,\mathcal D\Phi~ \e^{-S[\Phi]}~ \mathcal O_1(\Phi)\cdots
\mathcal O_n(\Phi) = \int\,\mathcal D(\Phi+\delta\Phi)~ 
\e^{-S[\Phi+\delta\Phi]}~ 
\mathcal O_1(\Phi+\delta\Phi)\cdots \mathcal O_n(\Phi+\delta\Phi)
\eeq
or equivalently to first order in $\delta\Phi_i$ one has the
Schwinger-Dyson equations
\beq
\int\,\mathcal D\Phi~\frac{\delta}{\delta\Phi_i} \left(
\e^{-S[\Phi]}~ \mathcal O_1(\Phi)\cdots \mathcal O_n(\Phi)\right) = 0
\ .
\eeq
One issue is whether or not the functional integration, or the path integral
measure, gives some additional contributions to these expressions. In
standard quantum field theory parlance, this is the same as the
question of whether or not there is some form of an anomaly in the
quantum theory.

To investigate this problem more precisely, we couple the fields
$\Phi_i$ to external source fields $J^i$ through the pairing
$(J,\Phi)\doteq\int\,\dd^2x~J^i\,\Phi_i$. The response of the system
to these external sources is encoded in the generating
functional for connected Green's functions given by
\beq
\mathcal{W}[J] = -\log\int \, \mathcal D\Phi~ \e^{-S[\Phi]-(J,\Phi)} \ .
\label{genfnal}\eeq
The effective action is defined by the Legendre transform
\beq
\Gamma\big[\,\hat\Phi\,\big] \doteq
\left(J^i\,\hat\Phi_i -\mathcal{W}[J]\right)_{\frac{\delta
    \mathcal{W}[J]}{\delta J^i}=\hat\Phi_i}
\ .
\eeq
Under the transformation (\ref{deltaPhi}), we make the natural
assumption that the functional integration measure is invariant, or
equivalently that in the path integral one can identify
\beq
\mathcal D(\Phi+\delta\Phi)~ \e^{-S[\Phi+\delta\Phi]} = 
\mathcal D\Phi~ \e^{-S[\Phi]} \ .
\label{deltameas}\eeq
Then we obtain the fundamental Ward identity
\beq
0 = \int\,\dd^2x~ \frac{\delta \Gamma\big[\,\hat\Phi\,\big]}
{\delta \hat\Phi{}_i}~ \frac{\langle0|\,\delta\hat\Phi{}_i\,|0
\rangle_{J_{\hat\Phi{}_i}}}{\langle0|0\rangle_{J_{\hat\Phi{}_i}}}
\label{effactionsym}\eeq
where
\beq
J_{\hat\Phi{}_i} = - \frac{\delta\Gamma\big[\,\hat\Phi\,\big]}
{\delta\hat\Phi{}_i}
\eeq
and the vacuum expectation values are taken in the original
quantum field theory coupled to the sources $J_{\hat\Phi{}_i}$. The
Ward identity (\ref{effactionsym}) is a statement of the symmetry of
the quantum effective action. These statements all implicitly assume
that an appropriate regularization of the Green's functions has been
specified which respects the symmetry (\ref{deltaPhi}).

\subsection{Implementation of Twisted Symmetries\label{TwistDiffGen}}

Let us now examine what features of the above analysis can be extended
to the case where the variation (\ref{deltaPhi}) represents twisted
symmetries of a noncommutative field theory, and in particular twisted
diffeomorphisms. In implementing these identities, crucial use is made
of the Leibnitz rule for the variational (or the functional
derivative) operator. In the twisted field theory the Leibnitz rule is
also twisted, giving an additional contribution to the variations of
star-products of fields, or equivalently of the twist element $\twist$
as in (\ref{Moyaltwisttransf}). If anomalous behaviour arises as
described above, then the hidden symmetry represented by twisted
diffeomorphisms cannot be implemented at the quantum level. We will
begin by deducing the most general possible twisted spacetime symmetry
of noncommutative field theory.

Let us start with a simple, explicit example of a noncommutative field
theory for illustration. Consider a real scalar field theory with action
of the form
\beq
S_V[\phi] = \frac12\,\int \, \dd^2x~\big(\phi\,\Box \phi + m^2\, \phi^2 +
  2V(\phi)\big) \ .
\label{SVphi}\eeq
Let us assume, as is customary in noncommutative field theory, that
the path integral measure is the standard Feynman measure $\mathcal
D\phi=\prod_{x\in\real^2}\,\dd\phi(x)$ of
commutative quantum field theory. This definition guarantees that when
$V=0$ the generating functional (\ref{genfnal}) is given by
$\mathcal{W}[J]=(J,\mathcal{C}J)$, where $\mathcal{C}$ is the free
propagator of the quantum field theory. It also agrees in this case
with the twist-covariant functional integral constructed
in~\cite{Grosse:2001pr}. Then the identification
(\ref{deltameas}) holds, and so we need only study the twisted
transformation of the action (\ref{SVphi}) under a shift of the field
$\phi$ by an infinitesimal diffeomorphism $\delta_\xi\phi$. For this,
we use the twisted covariance of the star-product
(\ref{eqn:gentransrule}) as explained in Section~\ref{TwistDiffs}.

To deduce the general structure of the variation, consider first
noncommutative $\phi^3$-theory with interaction potential $V(\phi)=g\,
\phi\star\phi\star\phi$. One then finds
\begin{equation}
\begin{split}
S_{\phi^3}[\phi] ~\mapsto~& S_{\phi^3}[\phi] - \, g\,\int\, \dd^2x~
\mu\Big(\mu_0\big(\delta_\xi\triangleright\mathcal F(\phi
\otimes\phi)\big) \otimes \phi\Big) - g\,\int\,
\dd^2x~\mu\Big(\phi\otimes\mu_0\big(\delta_\xi\triangleright
\mathcal F(\phi \otimes\phi)\big)\Big) \\[4pt]
~=~& S[\phi] -  \, 2  g\,\int\,
\dd^2x~\mu_0\big(\delta_\xi\triangleright
\mathcal F(\phi \otimes\phi)\big) \star \phi \ ,
\end{split}
\end{equation}
where we have used the trace property
\beq
\int\,\dd^2x~f_1\star f_2~=~\int\,\dd^2x~f_1\, f_2~=~\int\,\dd^2x~
f_2\star f_1
\label{traceprop}\eeq
for any two Schwartz fields $f_1,f_2\in\alg$.
For noncommutative $\phi^4$-theory with interaction potential
$V(\phi)=g\,\phi\star\phi\star\phi\star\phi$, one instead has
\begin{equation}
\begin{split}
S_{\phi^4}[\phi]~\mapsto S_{\phi^4}[\phi]\, - &\, g\,\int\, \dd^2x~
\mu\Big[\mu\Big(\mu_0\big(\delta_\xi\triangleright\mathcal F(\phi
\otimes\phi) \big) \otimes \phi\Big)\otimes\phi\Big] \\
- &\, g\,\int\, \dd^2x~\mu\Big[\mu\Big(\phi\otimes\mu_0\big(
\delta_\xi\triangleright\mathcal F(\phi
\otimes\phi)\big)\Big)\otimes\phi\Big] \\
- &\, g\,\int\, \dd^2x~\mu\Big(\mu(\phi\otimes\phi)\otimes\mu_0
\big(\delta_\xi\triangleright\mathcal F(\phi \otimes\phi)\big)\Big)
\\[4pt] = ~S[\phi]\, - & \, 3 g\, \int\, \dd^2x~\mu_0\big(
\delta_\xi\triangleright\mathcal F(\phi \otimes\phi)\big)
\star\phi\star\phi \ .
\end{split}
\end{equation}
These formulas generalize to arbitrary polynomial interactions in the
obvious way.

We see that the general structure of the variational terms that we
need to manipulate are of the form
\begin{equation}
\begin{split}
\int \, \dd^2x~\mu_0\big(\delta_\xi\triangleright\mathcal F(f_1\otimes
f_2)\big) = - \int \, \dd^2x~\sum_{k=1}^\infty\,
\frac{({-\ii\theta}/{2})^k}{k!}~\sum_{l=0}^k&\,\binom{k}{l}\,(-1)^l\,
\bigg[\big(\delta^{(R_1)}_{\partial_1^{k-l}\partial_2^l\xi}f_1\big) \, 
\big(\partial_2^{k-l}\partial_1^l f_2\big) \\
& + (-1)^k \,\big(\partial_2^{k-l}\partial_1^l f_1\big) \,
\big(\delta^{(R_2)}_{\partial_1^{k-l}\partial_2^l\xi}f_2\big)\bigg] \ .
\end{split}
\label{genvarterms}\end{equation}
When $f_1$ and $f_2$ are scalar Schwartz fields on $\real^2$, after
integrating by parts this becomes
\begin{equation}
\begin{split}
\int \,& \dd^2x~\mu_0\big(\delta_\xi\triangleright
\mathcal F(f_1\otimes f_2)\big)\\[4pt]
&~=~ - \int \, \dd^2x~\sum_{k=1}^\infty\, \frac{({-\ii\theta}/
{2})^k}{k!}~\sum_{l=0}^k\,\binom{k}{l}\,(-1)^l \,\bigg[
\big(\partial_1^{k-l}\partial_2^l\xi^\mu\big)\,\big(\partial_2^{k-l}
\partial_1^l f_2\big)\, \big(\partial_\mu f_1\big) \\
& \hspace{8cm}+ (-1)^k\, \big(\partial_2^{k-l}\partial_1^l f_1\big) \,
\big(\partial_1^{k-l}\partial_2^l\xi^\mu\big)\,
\big(\partial_\mu f_2\big)\bigg]\\[4pt]
& ~=~ - \int \, \dd^2x~\sum_{k=1}^\infty\, \frac{({\ii\theta}/
{2})^k}{k!}~ \sum_{l=0}^k\,\binom{k}{l}\,(-1)^l \,\xi^\mu \, 
\partial_\mu \bigg[ \big( \partial_1^{k-l}\partial_2^l f_1\big)\,
\big(\partial_2^{k-l}\partial_1^l f_2\big) \\ & \hspace{9cm}  + 
\big(\partial_1^{k-l}\partial_2^l f_1\big) \,
\big(\partial_2^{k-l}\partial_1^l f_2\big)\bigg] \ .
\end{split}
\end{equation}
After another integration by parts, the integral vanishes when
\beq
{\rm div}(\xi)=\partial_\mu\xi^\mu = 0 \ .
\label{divxi0}\eeq
This is a necessary and sufficient condition for the vector field
$\xi$ to generate an infinitesimal area-preserving diffeomorphism of
$\real^2$. It follows that noncommutative scalar field theory
possesses twisted symplectic symmetry.

This calculation can be extended to cases in which $f_1$ and $f_2$ are
not scalar fields. Consider as an example the contraction
$V_\sigma\star W^\sigma$. Then the part of (\ref{genvarterms})
corresponding to the transformations with respect to contractions of
the indices is given by the integral
\begin{equation}
\begin{split}
- \int \,&  \dd^2x~\sum_{k=1}^\infty
\,\frac{({-\ii\theta}/{2})^k}{k!}~\sum_{l=0}^k\,\binom{k}{l}\,
(-1)^l \,\bigg[ \big(\partial_1^{k-l}\partial_2^l\partial_\mu
\xi^\sigma\big)\,\big(\partial_2^{k-l}\partial_1^l
V_\sigma\big)\,W^\mu \\
& \hspace{8cm}- (-1)^k\, \big(\partial_2^{k-l}\partial_1^l 
W^\sigma\big) \,\big(\partial_1^{k-l}\partial_2^l
\partial_\sigma\xi^\mu\big)\,V_\mu\bigg] \\[4pt]
& ~=~ - \int \, \dd^2x~\sum_{k=1}^\infty\,
\frac{({\ii\theta}/{2})^k}{k!}~\sum_{l=0}^k\,\binom{k}{l}\,(-1)^l \,
\bigg[ \partial_\mu\xi^\sigma\, \big( \partial_1^{k-l}\partial_2^l 
W^\mu\big)\,\big(\partial_2^{k-l}\partial_1^l V_\sigma\big) \\
& \hspace{8cm}-
\partial_\sigma\xi^\mu\,\big(\partial_1^{k-l}\partial_2^l 
W^\sigma\big) \,\big(\partial_2^{k-l}\partial_1^l V_\mu\big)\bigg] \ .
\end{split}
\end{equation}
The integral trivially vanishes after relabelling indices. This purely
algebraic fact is linked to the twisted covariance of the classical
field theory that we established in Section~\ref{TwistDiffs}, and it
enables us to extend the classical symplectic symmetry to twisted
noncommutative field theories involving fields of higher rank.

Thus given any noncommutative field theory which is classically
twist invariant under area-preserving diffeomorphisms, one can still
write down, modulo boundary contributions, the Ward identity
(\ref{effactionsym}) for the effective action $\Gamma$ with
$\delta\Phi_i$ an area-preserving diffeomorphism. This is just the
statement that the quantum effective action is twist invariant under
symplectic transformations. Recall that this holds {\em only if}
the condition (\ref{deltameas}) is satisfied. The twist invariance of
the standard Feynman path integral measure $\mathcal{D}\Phi$ has been
somewhat of a matter of debate in the
literature~\cite{Balachandran:2006pi,Tureanu:2006pb,Zahn:2006wt}. We
will not enter into this debate nor elaborate here on the construction
of the functional integral which defines the twisted quantum field
theory~\cite{Grosse:2001pr,Oeckl:1999zu}. Our main analysis of the
Wilson loop in the next section will be independent of this definition
in any case.

\newsection{Wilson Loops in Twisted Noncommutative Gauge
  Theory\label{WilsonTNGT}}

In this final section we will apply the formalism developed thus far
to investigate the twisted symmetries of Wilson loop averages in
two-dimensional noncommutative Yang-Mills theory. We will first
describe the construction of Wilson loop operators in the twisted
quantum gauge theory, and then proceed to analyse the correlators. We
will find that the correlation functions of the usual noncommutative
Wilson loop operators are {\it not} twist invariant
under area-preserving diffeomorphisms of $\real^2$. Contrary to the
untwisted case, where even the classical gauge theory is only
invariant under the global $\mathfrak{sl}(2,\real)\rtimes\real^2$
subalgebra described in Section~\ref{APD}~\cite{Bassetto:2005hn}, the
Wilson loop operators on their own break the full classical
$W_{1+\infty}$ symmetry of the twisted gauge theory at the quantum
level (up to global symplectic transformations). This provides a
rather more general, explicit and nonperturbative description for this
loss of invariance than those given previously
in~\cite{Ambjorn:2004ck,Bassetto:2005hn,Cirafici:2005af}.

\subsection{Wilson Loop Operators\label{WilsonOps}}

Wilson loops are defined in terms of the holonomy operator
\beq
\hol^{(R)} \,:\, \Omega\mathcal M ~\longrightarrow~ G^{(R)} \ ,
\eeq
where $\Omega\mathcal M$ is the loop space of $\mathcal{M}=\real^2$,
$G$ is the gauge group, and $R$ is an irreducible unitary
representation of $G$. Let us examine the effect of the twist
deformation on this operator. The twist changes the way in which the
group $G$ (and therefore $\hol^{(R)}(\gamma)$ for each
$\gamma\in\Omega\mathcal{M}$) acts on tensor products of fields, but
not the group structure itself. This implies that the composition law
on the group of gauge transformations is the ordinary pointwise one
induced by the multiplication in $G$, i.e. if $u_1,u_2:\mathcal{M}\to
G$ are gauge transformations, then $u_1\circ u_2\doteq u_1\,u_2$. In
particular, there is no restriction on the allowed gauge groups for
which the twist deformation can be
implemented~\cite{Aschieri:2006ye,Vassilevich:2006tc}. This is
radically different from the usual definition in noncommutative gauge
theory~\cite{DougNek1,Sz1}, and has also been a focal point of much
debate in the
literature~\cite{Alvarez-Gaume:2006bn,Chaichian:2006we,Chaichian:2006wt}.

Let us regard a given loop $\gamma\in\Omega\mathcal M$ as a smooth
embedding
\beq
\gamma\,:\,[0,1]~\throw~ \mathcal M \ , \qquad s ~\mapsto~  z(s) 
\label{loopembed}\eeq
with $ z(0)= z(1)$. Let $A$ be a $\mfg$-valued connection one-form
on $\mathcal{M}$. Then we can define the holonomy operator
infinitesimally along $\gamma$ as
\begin{equation}
\hol^{(R)}(\gamma) = \sideset{^{\mathcal P}}{}\prod_{ z\in\gamma}\,
\big(\id^{(R)} + \ii~\dd z^\mu~ A^{(R)}_\mu( z)\big) \ ,
\label{eqn:holop}
\end{equation}
where the superscript $\mathcal{P}$ denotes path ordering along the
loop $\gamma$. In this setting it naturally corresponds to the
solution of the parallel transport equation
\beq
\frac{\partial\, \hol^{(R)}}{\partial s} + \dot z^\mu(s)\, 
A^{(R)}_\mu\big( z(s)\big)~ \hol^{(R)} = 0
\label{partranspeq}\eeq
where $\dot z^\mu(s)=\dd z^\mu(s)/\dd s$.

One can exponentiate the operator \pref{eqn:holop} to write
\beq
\hol^{(R)}(\gamma) = \mathcal P\, \exp\Big(\ii\oint_\gamma \,A^{(R)}\Big) \ .
\label{pathordexp}\eeq
The equivalence of the two formulas is provided by rewriting the
argument of the exponential in terms of the pullback under the map
(\ref{loopembed}) as
\beq
\oint_\gamma\, A = \int_{[0,1]}\, \gamma^*(A) \ ,
\eeq
from which we may expand the path-ordered exponential
(\ref{pathordexp}) explicitly as
\bea
\hol^{(R)}(\gamma) &=& 1 + \sum_{n=1}^{\infty}\, \ii^n~\int_{0}^{1}\,
\mathrm{d}s_1~\int_{0}^{s_1}\, \mathrm{d}s_2~\cdots
\int_{0}^{s_{n-1}}\, \mathrm{d}s_{n}~\dot z^{\mu_1}(s_1)\,
\dot z^{\mu_2}(s_2) \cdots \dot z^{\mu_n}(s_n)\nonumber\\ && 
\hspace{4cm}\times\, {A}^{(R)}_{\mu_1} \bigl( z(s_1)\bigr)\,
{A}^{(R)}_{\mu_2} \bigl( z(s_2)\bigr) \cdots
{A}^{(R)}_{\mu_n} \bigl( z(s_n)\bigr) \ .
\label{holRgammaexpand}\eea
Since the composition law of the group is not affected
by the twist, and since the loop (\ref{loopembed}) is an embedding of
the one-dimensional manifold $\mathbb{S}^1$ in $\mathcal M$, there are
no star-products required in the definition of the holonomy
operator. It thus transforms as usual under the ordinary adjoint
action of the gauge group.

This is very natural from an algebraic
point of view, as the gauge fields in (\ref{holRgammaexpand}) are
generically multiplied together at separated points and the holonomy
can thus be regarded as an element of the tensor algebra
\beq
\hol^{(R)}(\gamma)~\in~\bigoplus_{n=0}^\infty\,\Big(
\alg\,\otimes\,U\big(\mfg^{(R)}\big)\Big)^{\otimes n} \ .
\label{holtensalg}\eeq
As a consequence, there is no reason {\it a priori} to compose
(\ref{holtensalg}) with the deformed product
$\mu:\alg\otimes\alg\to\alg$. Since the holonomy operator is a
generator element of the group algebra $\hil=\complex G\supset G$, it
acts on tensor products of fields through the twisted coproduct as
\beq
\Delta\big(\hol^{(R)}(\gamma)\big) = \mathcal F^{-1}\circ
\big(\hol^{(R_1)}(\gamma) \otimes \hol^{(R_2)}(\gamma)\big)\circ
\mathcal F  \ .
\eeq
From the remarks made at the end of Section~\ref{TwistDiffs} it
follows that $\hol^{(R)}(\gamma)$ is classically covariant under
area-preserving diffeomorphisms, as well as being gauge covariant. 

However, this is not the only definition of Wilson loop operators
which leads to gauge invariant observables that are covariant under
area-preserving diffeomorphisms. We can systematically deform the
holonomy operators (\ref{holRgammaexpand}) into the star-holonomy
operators defined by
\bea
\mathcal P\, \exp_\star\Big(\ii\oint_\gamma \,A^{(R)}\Big) 
&\doteq& 1 + \sum_{n=1}^{\infty}\, \ii^n~\int_{0}^{1}\,
\mathrm{d}s_1~\int_{0}^{s_1}\, \mathrm{d}s_2~\cdots
\int_{0}^{s_{n-1}}\, \mathrm{d}s_{n}~\dot z^{\mu_1}(s_1)\,
\dot z^{\mu_2}(s_2) \cdots \dot z^{\mu_n}(s_n)\nonumber\\ && 
\hspace{2cm}\times\, {A}^{(R)}_{\mu_1} \bigl( z(s_1)\bigr)
\star{A}^{(R)}_{\mu_2} \bigl( z(s_2)\bigr) \star\cdots\star
{A}^{(R)}_{\mu_n} \bigl( z(s_n)\bigr) \ .
\label{holstar}\eea
This definition amounts to replacing the tensor product algebra in
(\ref{holtensalg}) with the {\it braided} tensor product
algebra~\cite{Aschieri:2005zs,Fiore:2007vg,Tureanu:2006pb}, as
naturally imposed by non-cocommutativity of the twisted coproduct
$\Delta$. From the general analysis of Section~\ref{TwistSym} it
follows that these operators possess both twisted gauge covariance and
twisted symplectic covariance. They are also invariant under the
adjoint action of the star-gauge group~\cite{DougNek1,Sz1}, which is
defined to be the gauge group $G$ with its multiplication deformed by
the star product, i.e. if $u_1,u_2:\mathcal{M}\to G$ are
gauge transformations, then $u_1\circ u_2\doteq u_1\star u_2$. In
contrast to the twist deformation, the star-product deformation limits
the choice of gauge group and of representation
$R$~\cite{Bonora:2000td,Chaichian:2001mu,Terashima:2000xq}, in the
simplest settings to the unitary groups $G={\rm U}(N)$ and to the
$N$-dimensional fundamental representation. On the other hand, twisted
gauge covariance of (\ref{holstar}) generically requires the gauge
connection $A_\mu$ to be $U(\mfg)$-valued~\cite{Aschieri:2006ye} so
that the star-holonomy is in general a non-generator element of the
group algebra $\complex G^{(R)}$.

The definition of the star-holonomy operator, given in (\ref{holstar})
as a star-deformation of the ordinary gauge holonomy in
(\ref{holRgammaexpand}), can be understood more fundamentally in the
context of twist deformations as follows. In the argument of the
path-ordered exponential (\ref{pathordexp}) we can regard the contour
integral as a bilinear map
\beq
\nu_0=\oint \,:\, \Omega\mathcal M \otimes{\mathcal
  V}\big(\mathcal{M}\,,\,U(\mfg)\big)  ~\throw~ U(\mfg) \ ,
\label{contintmap}\eeq
where ${\mathcal V}(\mathcal{M},U(\mfg))$ is the affine space of
enveloping algebra valued connection one-forms on $\mathcal{M}$ and
the linearity with respect to the first factor is defined in terms of
the natural $\mathbb Z$-module structure on the loop space
$\Omega\mathcal{M}$. All three of the linear spaces in
(\ref{contintmap}) carry a natural action of the semi-direct product
of the group of gauge transformations with the diffeomorphism group of
$\mathcal{M}$. Then just as we defined the twisted product in
(\ref{twistalg}), following the philosophy of~\cite{Aschieri:2005zs}
we can deform the map (\ref{contintmap}) by
combining it with an action of the twist to get
$\nu\doteq\nu_0\circ\twist$. The twisted map $\nu$ satisfies the usual
covariance condition $g\circ\nu=\nu\circ\Delta(g)$. This is a
conceptually useful point of view for calculational purposes, as it
allows one to naturally combine variations of the gauge connection
$A_\mu$ with reparametrizations of the contour $\gamma$ (i.e. of the
embedding $ z^\mu(s)$) in an invariant way.

The star-holonomy operators obey a natural star-deformed version of
the parallel transport equation
(\ref{partranspeq})~\cite{Alekseev:2000td,Sz1}. The corresponding
Wilson loop is denoted
\beq
W_\star[\gamma,A]\doteq \Tr\,\mathcal P\,
\exp_\star\Big(\ii\oint_\gamma \,A\Big) \ ,
\label{starWilsondef}\eeq
where for brevity we drop the representation label from the
notation. Although the ordinary holonomy operators $\hol^{(R)}(\gamma)$
yield the natural observables with respect to the twisted gauge
symmetry, in the remainder of this paper we will focus on star-gauge
invariant observables constructed from the star-Wilson
loops~(\ref{starWilsondef}). The distinction between the two classes
of observables is similar to the distinction between deformed and
undeformed products (or braided and unbraided tensor products) of
fields at separated spacetime
points~\cite{Balachandran:2006pi,Fiore:2007vg,Tureanu:2006pb}. In some
instances, the incorporation of both types of observables reflects
invariance of the noncommutative gauge theory under both twisted and
star gauge symmetries, which has recently been argued to be a generic
requirement for consistency of twisted gauge
theories~\cite{Giller:2007gq}. Detailed comparisons between the two
types of gauge symmetries can be found in~\cite{Banerjee:2006jy}.

\subsection{Star-Gauge Invariant Correlators\label{WilsonCorrs}}

Noncommutative Yang-Mills theory on $\real^2$ is defined by the action
functional
\beq
S_{\rm YM}^\star[A]=\frac1{2}\,\int\,\dd^2x~\Tr f^2 \, 
\label{SNCYMdef}\eeq
where $f=F_A=\partial_1A_2-\partial_2A_1-\ii e\,(A_1\star A_2-A_2\star
A_1)$ is the $U(\mfg)$-valued noncommutative field strength. As the
action (\ref{SNCYMdef}) is defined by the star product, it is natural
to consider star-gauge invariant correlation functions of the form
\beq
\int\,\mathcal DA~ \e^{-S_{\rm YM}^\star[A]}~ 
\mathcal O_1^\star(A)\cdots \mathcal O_n^\star(A)
\label{NCYMgencorrs}\eeq
where the operators $\mathcal O_i^\star(A)$ include the star-Wilson loops
(\ref{starWilsondef}). The ordinary, undeformed Yang-Mills action
$S_{\rm YM}[A]$ is invariant under area-preserving diffeomorphisms of
$\real^2$, because in two dimensions the field strength $f$ is a
scalar field and hence any diffeomorphism which preserves the area
form $\dd^2x$ is a symmetry of the action. Furthermore, any Wilson
loop, constructed from a holonomy operator $\hol^{(R)}(\gamma)$,
is a homotopy invariant which depends only on the area enclosed by the
loop $\gamma$, and not on the shape of $\gamma$. This classical
symmetry extends to the quantum gauge theory~\cite{Witten:1992xu}, and
in particular the functional integral measure
$\mathcal{D}A$ is invariant under the corresponding transformations of
the gauge connection $A$. From the general analysis of Section~\ref{TwistSym}
(see remarks at the end of Section~\ref{TwistDiffs}), it follows that
the classical noncommutative gauge theory is thus invariant under
twisted area-preserving diffeomorphisms. We will now study the
behaviour of the correlators (\ref{NCYMgencorrs}) under the twisted
symplectic symmetry, and decide whether or not it is
preserved after quantization.

To work out Ward identities in path integral quantization for this
case, we proceed as in Section~\ref{TwistNCFT} to first determine whether or
not the correlation functions exhibit anomalous breakdown of the
classical symplectic symmetry. For this, we expand the gauge fields as
$A_\mu=A_\mu^a~T^a$ where $T^a$ are the generators of the gauge group
with $\Tr(T^a\,T^b)=\delta^{ab}$, and couple them to external currents
$J^\mu=J^\mu_a~T^a$ through the pairing
$(J,A)\doteq\int\,\dd^2x~J^\mu_a\,A_\mu^a$. Then the general arguments of
Section~\ref{TwistNCFT} can be used to infer that the quantum
effective action constructed through a Legendre transformation of the
connected generating functional
\beq
-\log\int\,\mathcal DA~ \e^{-S_{\rm YM}^\star[A]-(J,A)}
\eeq
is invariant under twisted area-preserving diffeomorphisms. A possible
obstruction to this reasoning in the present case is the implicit
presence of the Faddeev-Popov determinant in the path integral measure
$\mathcal{D}A$. Since it is constructed from the Gauss' law
constraint, it is built by means of star-products of the gauge
fields and hence we need to show that it is twist invariant under a
variation of the gauge field $A$ by an area-preserving
diffeomorphism.

That this is indeed the case can be seen by expressing the determinant
as a functional integral over anticommuting ghost fields $\bar
c\,^a,c^a$ in the adjoint representation of the gauge group with the
Lorentz gauge action
\bea
S_{\rm gh}^\star[\,\bar c,c,A] &=& 
\int\, \dd^2x ~\left( \bar c\,^a \,\square c^a + e\,
  f^{abh} \,\bar c\,^a \star A_\mu^{ b} \star\partial^\mu c^h
\right.\nonumber\\ && \hspace{2cm}- \left.
\ii e \,d^{abh}\,\bar c\,^a \star \big( A^{h}_\mu \star \partial^\mu 
c^b-\partial^\mu c^b\star A^{h}_\mu\big)\right) \ ,
\label{Sgh}\eea
where $f^{abh}$ are the structure constants of the Lie algebra $\mfg$
and $d^{abh}$ are invariant tensors for the
adjoint action of $\mfg$. The proof of twisted invariance of integrals
over scalar densities given in Section~\ref{TwistDiffGen} can now be
applied to (\ref{Sgh}). We conclude that the twisted symplectic
symmetry is non-anomalous in this case. Hence we need only further
study the behaviour of the star-Wilson loops (\ref{starWilsondef})
under twisted area-preserving diffeomorphisms, implemented at the
quantum level in (\ref{NCYMgencorrs}).

\subsection{Twist Transformations of Quantum Star-Wilson
  Loops\label{WilsonTwist}}

To analyse the twisted quantum symmetries of the operators
(\ref{starWilsondef}), we shall use a proper path integral
representation. For this, we introduce a pair of
one-dimensional complex auxilliary fields $\bar\zeta_i(s)$ and
$\zeta_i(s)$ on the loop $\gamma$ which transform respectively in the
fundamental and anti-fundamental representations of the gauge
group. Their propagator is given by
\beq
\big\langle \bar\zeta_i(s_1)\, \zeta_j(s_2) \big\rangle_\zeta
\doteq \int \,\mathcal D \bar \zeta~ \mathcal D \zeta~
\bar\zeta_i(s_1)\,\zeta_j(s_2)~ \exp\Big(-\int_0^1\, \dd
s~\bar\zeta_k(s-0^+)\,\dot\zeta_k(s)\Big)= \delta_{ij}~
\Theta(s_1-s_2) \nonumber\\
\label{auxprop}\eeq
where $\Theta(s)$ is the Heaviside step function. The regularization
indicated in the functional integral takes care of the ambiguous value
of $\Theta(s)$ at $s=0$. The technical details of this regularization
play no role below. They are analysed in~\cite{Mavromatos:1998nz}
(see also~\cite{Ambjorn:2004ck} for an analysis in a slightly
different context).

By using the propagator (\ref{auxprop}) and Wick's theorem, we can
unravel the path ordering operation
$\mathcal{P}$~\cite{Mavromatos:1998nz} in
(\ref{starWilsondef}) to write the star-Wilson loop operator as
\begin{equation}
W_\star [\gamma,A]= \Big\langle \bar\zeta_k(0) ~\exp_\star\Big(
\ii\int_0^1 \,\dd s~\bar\zeta_i(s-0^+) \,A^{ij}_\mu \big(z(s)\big)\, 
\zeta_j(s)\,\dot z^\mu(s)\Big)~ \zeta_k(1)\Big\rangle_\zeta \ .
\label{eqn:wilunwrap}
\end{equation}
We have dropped an irrelevant factor induced by the vacuum graphs of
the auxilliary quantum field theory on $\gamma$ (which is equal to one for
unimodular gauge groups and can be handled in exactly the same way as
we do with (\ref{eqn:wilunwrap}) below). In (\ref{eqn:wilunwrap}) the
star-products act only on the arguments of the gauge fields and do not
involve the auxilliary fields. To avoid cluttered formulas below, we
therefore introduce the shorthand notation 
\beq
\underline{A}\,_\mu\big(z(s)\big) \doteq 
\bar\zeta_i(s-0^+)\,A^{ij}_\mu\big(z(s)\big)\,\zeta_j(s) \ .
\label{shorthand}\eeq
These fields can be formally star multiplied in (\ref{eqn:wilunwrap})
without worrying about reordering each term, as our manipulations will
not depend on the details of the loop integrals anyway.

We now expand the exponential in (\ref{eqn:wilunwrap}) and check for
twisted symplectic invariance order by order in the loop embedding
functions $z^\mu$. The zeroth and first order terms are trivially
invariant as they do not involve any star products. The second order
term is proportional to
\beq
\int_0^1\,\dd s_1 ~ \int_0^1\, \dd s_2 ~
\underline{A}\,_\mu\big(z(s_1)\big)\,\star\, \underline{A}\,_\nu
\big(z(s_2)\big)\, \dot z^\mu(s_1)\,\dot z^\nu(s_2) \ .
\label{2ndorder}\eeq
Since there is no star product between $\underline{A}\,_\mu$ and $\dot
z^\mu$, the arguments used in Section~\ref{TwistDiffs} do not
guarantee the scalar property of this term and we must check its
twisted transformation properties directly. Consider the variation
\beq
\underline{A}\,_\mu ~\mapsto~ \underline{A}\,_\mu + 
\delta_\xi^{(R_1)} \underline{A}\,_\mu
\eeq
of the abelianized gauge field (\ref{shorthand}) under an
infinitesimal area-preserving diffeomorphism generated by a vector
field $\xi$, where $R_1$ is the representation of the diffeomorphism
group defined in (\ref{deltaxirank1}). Discarding that part which can
be absorbed into a deformation of the contour $\gamma$ under the
diffeomorphism, one finds that the variation of (\ref{2ndorder}) is
given by
\begin{equation}
\begin{split}
\int_0^1\, \dd s_1 ~ & \int_0^1\, \dd s_2 ~ \dot z^\mu(s_1)\,
\dot z^\nu(s_2) \\ & \times~
\sum_{k=1}^\infty\, \frac{({-\ii\theta}/{2})^k}{k!}~ 
\sum_{l=0}^k\, \binom{k}{l}\,(-1)^l\, \bigg(
\delta^{(R_1)}_{\partial_1^{k-l}
\partial_2^l\xi}\, \underline{A}\,_\mu\big(z(s_1)\big) 
\star\partial_2^{k-l}
\partial_1^l\, \underline{A}\,_\nu\big(z(s_2)\big) \\ & \hspace{4cm}
+ (-1)^k\, \partial_2^{k-l}\partial_1^l\, 
\underline{A}\,_\mu\big(z(s_1)\big) 
\star \delta^{(R_1)}_{\partial_1^{k-l}\partial_2^l\xi}\, 
\underline{A}\,_\nu\big(z(s_2)\big) \bigg) \ .
\end{split}
\label{eqn:whole}
\end{equation}

Each term of the summations in (\ref{eqn:whole}) is of the generic
form
\begin{equation}
\begin{split}
\int_0^1 \,\dd s_2 ~ & \dot z^\nu(s_2)~ \int_0^1 \,\dd s_1 ~ 
\dot z^\mu(s_1) \\ & \times\,\Big(
\partial_1^{k-l}\partial_2^l\xi^\sigma\,\partial_\sigma \,
\underline{A}\,_\mu\big(z(s_1)\big)+\partial_\mu\partial_1^{k-l}
\partial_2^l\xi^\sigma\, \underline{A}\,_\sigma\big(z(s_1)\big)\Big) 
\star\partial_2^{k-l}\partial_1^l\,
\underline{A}\,_\nu\big(z(s_2)\big) \ .
\end{split}
\end{equation}
At this point, it is tempting to use the twist deformation of the
bilinear contour map (\ref{contintmap}) described in
Section~\ref{WilsonOps} above to absorb the variation of the
contravariant vector field $\underline{A}\,_\mu$ into a deformation of
the loop $\gamma$ (by appropriately redefining the embedding
functions $z^\mu(s)$). But this would produce a sum of loop integrals,
each one taken over a different contour. The only straightforward way
to proceed is to restrict ourselves to some subgroup of the group of
area-preserving diffeomorphisms and study its contribution to
(\ref{eqn:whole}).

Let us expand the smooth vector field $\xi$ in a Taylor series
\beq
\xi^\mu(x) \doteq \sum_{l=0}^\infty\,L^\mu{}_{\sigma_1\cdots\sigma_l}~
x^{\sigma_1} \cdots x^{\sigma_l}
\label{APDexpand}\eeq
where $L^\mu{}_{\sigma_1\cdots\sigma_l}$ is a Schwartz sequence of
constant tensors which are completely symmetric in their lower
indices. The divergence free constraint (\ref{divxi0}) is equivalent
to the traceless conditions
\beq
L^\mu{}_{\sigma_1\cdots\sigma_l\mu}=0 \ .
\label{traceless}\eeq
The expression \pref{eqn:whole} may then be rewritten in the
convenient form
\begin{equation}
\begin{split}
&\int_0^1\, \dd s_1 ~  \int_0^1 \,\dd s_2 ~ \dot z^\mu(s_1)\, 
\dot z^\nu(s_2) ~ \sum_{l=0}^\infty \,\bigg( z^{\alpha_1}(s_1)\cdots 
z^{\alpha_l}(s_1) ~\sum_{k=1}^\infty\,\frac{\ii^k}{k!}\,
\theta^{\sigma_1\rho_1}\cdots\theta^{\sigma_k\rho_k} \\ &
\times\,
\Big(L^\rho{}_{\mu\sigma_1\cdots\sigma_k\alpha_1\cdots\alpha_l}\,
\underline{A}\,_{\rho}\big(z(s_1)\big)+
\delta^\rho{}_\mu
\,L^\sigma{}_{\sigma_1\cdots\sigma_k\alpha_1\cdots\alpha_l}\,
\partial_\sigma\,\underline{A}\,_{\rho}\big(z(s_1)\big)\Big)\, \star
\,\partial_{\rho_1}\cdots\partial_{\rho_k}\,
\underline{A}\,_{\nu}\big(z(s_2)\big) \\ & \hspace{4cm} + 
\textrm{ sym. } \bigg)
\end{split}
\label{convenient}\end{equation}
where the only non-vanishing components of the tensor
$\theta^{\mu\nu}$ are $\theta^{12}=-\theta^{21}=\theta$, and the
additional term is the corresponding symmetric contribution with $s_1$
and $s_2$ interchanged.

Let us first examine the expression (\ref{convenient}) in the case of
a unimodular linear affine transformation (\ref{affine}), i.e. a
global area-preserving diffeomorphism in
$\mathfrak{sl}(2,\real)\rtimes\real^2$. One obtains
\begin{equation}
\begin{split}
\int_0^1 \,\dd s_1 ~ \int_0^1\, \dd s_2 ~ \dot z^\mu(s_1) \,
\dot z^\nu(s_2) \ii \theta^{\sigma_1\rho_1}\,\Big( & \delta^\rho{}_\mu\,
L^\sigma{}_{\sigma_1}~\partial_\sigma\,
\underline{A}\,_{\rho}\big(z(s_1)\big)\star \partial_{\rho_1}\,
\underline{A}\,_{\nu}\big(z(s_2)\big) \\ 
& + \delta^\rho{}_\nu\,L^\sigma{}_{\rho_1}  ~\partial_{\sigma_1}\,
\underline{A}\,_{\mu}\big(z(s_1)\big)\star\partial_{\sigma}\,
\underline{A}\,_{\rho}\big(z(s_2)\big)\Big) \\[4pt]
~= ~\int_0^1 \,\dd s_1 ~ \int_0^1 \,\dd s_2 ~ \dot z^\mu(s_1)\,
\dot z^\nu(s_2) &
\ii\theta^{\sigma_1\rho_1}\,\big(L^\sigma{}_{\sigma_1}\,
\delta^\rho{}_{\rho_1} + \delta^\sigma{}_{\sigma_1}\,
L^\rho{}_{\rho_1}\big) \\ & \hspace{1cm}
\times~\partial_\sigma\, \underline{A}\,_{\mu}
\big(z(s_1)\big)\star \partial_\rho\,\underline{A}\,_{\nu}
\big(z(s_2)\big) \ .
\end{split}
\label{linAPDterm}\end{equation}
Using the traceless property (\ref{traceless}) one sees that
(\ref{linAPDterm}) vanishes. However, with a bit of work it can be
seen that quadratic and higher-order terms in $x^\mu$ in the expansion
(\ref{APDexpand}) of the vector field $\xi$ do not vanish even with
the traceless property (\ref{traceless}) taken into account. The
generalization of this calculation to arbitrarily high orders in the
expansion of the exponential in \pref{eqn:wilunwrap} is
straightforward.

We conclude that the star-Wilson loop is twist invariant only under
{\em global} area-preserving diffeomorphims, similarly to the property
that linear affine transformations are the only spacetime symmetries
under which the Moyal product is
covariant~\cite{Gracia-Bondia:2006yj}. However, this {\em does not}
mean that the star-Wilson loop is not invariant under general twisted
symplectic transformations, but only that our attempt to realize
twisted area-preserving diffeomorphisms as a symmetry of the
noncommutative quantum field theory cannot be implemented on the
star-Wilson loop in a way that keeps it invariant. This result
explains the shape dependence and loss of non-linear symplectic
invariance of the correlators of Wilson loops observed previously
in~\cite{Ambjorn:2004ck,Bassetto:2005hn,Cirafici:2005af}. It
also applies to any operator built from the star-Wilson loop, and it
persists for open Wilson lines as our arguments hold as well when
the contour $\gamma$ is not closed. Thus any star-gauge invariant
observable will break the twisted symmetry under non-linear area-preserving
maps. Since the effect of the twist acts on the contour $\gamma$
globally, we also see that the twist variation cannot be cancelled by
means of {\em any} local operator insertion. The only remaining hope
for full symplectic invariance is an alternative definition of the
twisted noncommutative quantum gauge theory using the braided Wick
expansions of~\cite{Oeckl:1999zu}, for example, which has recently
been argued to be the correct framework for the implementation of
twisted symmetries in noncommutative field
theory~\cite{Balachandran:2006pi}.

\subsection*{Acknowledgments}

We thank A.P.~Balachandran, C.~S\"amann, H.~Steinacker, S.~Vaidya and
J.D.~Vergara for helpful discussions. This work was supported in part
by the EU-RTN Network Grant MRTN-CT-2004-005104. The work of M.R. was
supported in part by a Fellowship of the {\sl Fondazione Angelo della
  Riccia}.

\end{document}